\documentclass[epj]{svjour}
\usepackage{graphics}
\usepackage{color}
\usepackage{bm}
\begin{document}
	\title{Gyrotactic cluster formation of bottom-heavy squirmers}
	\author{Felix R\"uhle\inst{1} \and Arne W. Zantop\inst{1} \and Holger Stark\inst{1}
	}                     
	%
	%
	\institute{Institut f\"ur Theoretische Physik, Technische Universit\"at Berlin, Hardenbergstr. 36, D-10623 Berlin, Berlin, Germany}
	\date{Received: date / Revised version: date}
	%
	\abstract{
		Squirmers that are bottom-heavy experience a torque that aligns them along the vertical so that they swim upwards. In a 
		suspension of many squirmers, they also interact hydrodynamically via flow fields that are initiated by their swimming motion 
		and by gravity. Swimming under the combined action of flow field vorticity and gravitational torque is called gyrotaxis. 
		Using the method of multi-particle collision dynamics, we perform hydrodynamic simulations of a many-squirmer system floating 
		above the bottom surface. Due to gyrotaxis they exhibit pronounced cluster formation with increasing gravitational torque.
		The clusters are more volatile at low  values but compactify to smaller clusters at larger torques. The mean distance between clusters is mainly controlled by the gravitational torque and not the global density. Furthermore, we 
		observe that neutral squirmers form clusters more easily, whereas pullers require larger gravitational  torques due to their additional 
		force-dipole flow fields. We do not observe clustering for pusher squirmers. Adding a rotlet dipole to the squirmer flow field induces
		swirling clusters.
		At high gravitational strengths, the hydrodynamic interactions with the no-slip boundary create an additional vertical alignment for neutral squirmers, which also supports cluster formation.
	} 
	\maketitle
	\section{Introduction}
	\label{intro}
	The collective dynamics of active entities features intriguing pattern formation on different length scales~\cite{Weihs1973,VicsekShochet1995,SaintillanShelley2008,CavagnaViale2010,JeckelDrescher2019}.
	It also offers the prospect of manufacturing new materials by aggregating synthetic agents~\cite{PeyerNelson2013,VachFaivre2017,BricardBartolo2013}. 
	Biotic cluster or patch formation of phytoplankton has profound influences on oceanic ecology~\cite{GeninFranks2005,DurhamStocker2009,BenziToschi2012}, while bacteria forming biofilms endanger 
	human health~\cite{WatnickKolter2000,StantonSanchez2017}.
	Investigating how active particles react on external fields reveals 
	novel aspects of these non-equilibrium systems~\cite{Stark2016,GompperKale2020}. This includes
	sedimentation of active particles~\cite{TailleurCates2009,PalacciBocquet2010,EnculescuStark2011,WolffStark2013,GinotCottin-Bizonne2015,RuehleStark2018}, 
	microswimmers in magnetic fields~\cite{CebersOzols2006,Guzman-LastraLoewen2016,FanSandoval2017},
	their pearling transition and plume formation in microchannels~\cite{WaisbordCottin-Bizonne2016,TheryLauga2020}, as well as the question of optimal steering of active entities under flow and in a potential landscape \cite{ColabreseBiferale2017,SchneiderStark2019,DaddiMoussaIderLiebchen2020}.
	
	In this article, we focus on microswimmers under gravity, which is of utmost importance for artificial and biological systems alike~\cite{TailleurCates2009,PalacciBocquet2010,EnculescuStark2011,GinotCottin-Bizonne2015,PedleyKessler1992,Roberts2006,CampbellEbbens2013,tenHagenBechinger2014}. 
	The non-equilibrium sedimentation of active particles has attracted a lot of attention~\cite{TailleurCates2009,PalacciBocquet2010,EnculescuStark2011,GinotCottin-Bizonne2015,KuhrStark2017}, 
	revealing an activity-dependent sedimentation length~\cite{PalacciBocquet2010}, polar order \cite{EnculescuStark2011}, 
	and convection~\cite{KuhrStark2017}. Under gravity often a gravitational torque acts on microswimmers due to their non-spherical shape or a non-uniform mass distribution, which is responsible for negative gravitaxis~\cite{WolffStark2013,Roberts2006,CampbellEbbens2013} 
	and plume formation~\cite{PedleyKessler1992,PlessetWinet1974,BeesHill1997,JanosiHorvath1998}.
	
	The collective motion of microswimmers 	is  profoundly influenced by hydrodynamic flow fields,
	which they create themselves
	\cite{SimhaRamaswamy2002,SaintillanShelley2011,AlarconPagonabarraga2013,ZoettlStark2014,BlaschkeStark2016,StenhammarMorozov2017}.
	Microswimmers residing at a bottom surface due to strong gravity  
	show hovering and trapped states~\cite{RuehleStark2018,UspalTasinkevych2015,SimmchenSanchez2016} as well as raft and swarm
	formation~\cite{KruegerMaass2016,KuhrStark2019},
	due to hydrodynamic
	wall interactions.
	Also external forces induce fluid flow, for example, when passive colloids sediment under gravity~\cite{CichockiHinsen1995}.
	An example from active matter are microswimmers in a harmonic trap.
	They experience novel pattern formation by creating a fluid pump \cite{NashCates2010,HennesStark2014}.
	
	Gyrotaxis refers to directed locomotion resulting from a combination of gravitational and viscous torques in a flow	\cite{Kessler1985ContPhys,DurhamStocker2011,Pedley2015}. 
	In particular, it applies to the action of fluid vorticities on microswimmers \cite{Kessler1985ContPhys,Pedley2015}.
	As such, it is responsible for the dancing motion of Volvox algae~\cite{DrescherGoldstein2009}, focussing in
	channel flow \cite{Kessler1985}, the formation of layers of phytoplankton~\cite{DurhamStocker2009}, and convective patterns in algae and bacteria~\cite{PedleyKessler1992,TimmOkubo1994,GhoraiHill1999,SokolovAranson2009}.
	These examples show the importance of gyrotaxis for biological systems.
	They motivated us, in an earlier article, to perfom hydrodynamic simulations using the methode of multi-particle collision dynamics 
	in order to to systematically study the dynamics and gyrotaxis of bottom-heavy squirmers. The latter are generic model microswimmers  \cite{RuehleStark2020}.
	Depending on the strength of gravity and bottom heaviness, we identified different dynamical states including inverted 
	sedimenation,  plumes and convective rolls, as well as spawning clusters.
	
	In this article we continue our work on gyrotaxis of bottom-heavy squirmers but now concentrate on moderate densities and strong gravity such that they float above but close to the bottom surface. Only with increasing gravitational torque gyrotactic cluster formation sets in as a subtle balance between hydrodynamic and gravitational torques. We thoroughly characterize the cluster formation of neutral squirmers for different areal densities and torques using different quantities such as the vertical density profile, mean cluster size, and radial distribution function. We also discuss the influence of squirmer flow fields from pusher and puller squirmers. While pusher squirmers do not exhibit noticeable cluster formation, lar\-ger torques are needed to observe it for puller squirmers. Furthermore, a hydrodynamic rotlet dipole realized,
	for example, by bacteria also weakens cluster formation.
	
	The article is structured as follows. In Sect.~\ref{sec:methods} we present the squirmer model and describe in detail the 
	hydrodynamics of this model swimmer under gravity. Furthermore, we introduce the method of multi-particle collision dynamics to 
	simulate the flow field generated by the squirmers. In Sect.~\ref{sec:results} we present our simulation results and we conclude in Sect.~\ref{sec:conclusions}.
	
	\section{Methods}  \label{sec:methods}
	
	\subsection{Squirmer model}
	
	We use the squirmer model introduced by Lighthill and Blake~\cite{Lighthill1952,Blake1971} in order to investigate gyrotactic cluster formation of microswimmers. 
	The original idea of the model
	is to consider a spherical particle with a surface velocity field and expand it into appropriate base function to model the surface actuation of a ciliated microswimmer. A more general ansatz 
	includes chiral surface flow and was presented in Ref.~\cite{PakLauga2014}, which we follow here. We 
	work with an axisymmetric tangential flow field.
	Then, the surface slip velocity for a squirmer with radius $R$ that swims along the unit vector $\mathbf{e}$ can be expressed as
	\begin{eqnarray}
		\label{eq:squirmer_model}
		\mathbf{u}_\mathrm{sq}(\mathbf{r})\vert_{r=R} & = & \sum_{n=1}^{\infty}B_n \frac{2 P^\prime_n(\mathbf{e}\cdot\mathbf{\hat{r}})}{n(n+1)}\left[ - \mathbf{e}  +  (\mathbf{e}\cdot\mathbf{\hat{r}}) \mathbf{\hat{r}} \right] \nonumber \\ 
		&  + &\sum_{n=1}^{\infty}C_n P^\prime_n(\mathbf{e}\cdot\mathbf{\hat{r}})\left[ \mathbf{e}  \times  \mathbf{\hat{r}}\right].
	\end{eqnarray}
	Here $P^\prime_n$ refers to the first derivative of the $n$th Legendre polynomial $P_n$.
	
	Often, one truncates this expansion after the second order. The mode $B_1$ is necessary for self-propulsion
	where  $v_0 = \frac{2}{3}B_1$ is the swimming velocity.
	Additionally, we  consider the modes $B_2$ and $C_2$, which we express in terms of the squirmer parameter $\beta = B_2/B_1$ 
	and chirality parameter $\chi = C_2/B_1$, see for example Ref.~\cite{FaddaYamamoto2020}. The parameter $\beta$ switches between 
	neutral squirmers ($\beta=0$), pullers ($\beta >0$) and pushers ($\beta < 0$).
	
	\subsubsection{Hydrodynamics of bottom-heavy squirmers}
	
	We collect all the flow and vorticity fields of a bottom-heavy squirmer and thereby provide an overview on the
	hydrodynamics of our squirmer system.
	
	\paragraph{Flow and vorticity fields}
	\label{sec:flowfields}
	The flow field of the freely swimming squirmer resulting from the truncated surface 
	field in eq.~(\ref{eq:squirmer_model})
	is~\cite{PakLauga2014}
	\begin{eqnarray}
		\label{eq:squirmer_field}
		\mathbf{u}_\mathrm{sq}(\mathbf{r}) & = & \frac{3}{2}v_0 \left[ \frac{1}{3}\frac{R^3}{r^3}\left(-\mathbf{\mathbf{e} + 3(\mathbf{e}\cdot\mathbf{\hat{r}})\mathbf{\hat{r}}}\right) \right. \nonumber \\ 
		& - &  \frac{\beta}{2}\frac{R^2}{r^2}\left( -1+ 3(\mathbf{e}\cdot\mathbf{\hat{r}})^2\right)\mathbf{\hat{r}} \\ 
		& + & \left. 3\chi \frac{R^3}{r^3} (\mathbf{e}\cdot\mathbf{\hat{r}}) \mathbf{e}  \times  \mathbf{\hat{r}}  + \mathcal{O}\left(\frac{R^4}{r^4}\right)
		\right] \nonumber,
	\end{eqnarray}
	where we have written the 
	different contributions 
	such that one immediately recognizes the source dipole (first term), force dipole (term with $\beta$) and rotlet dipole (term with $\chi$).
	
	Equation~(\ref{eq:squirmer_field}) shows that a squirmer always creates a source dipole field, which becomes the leading order for the 
	neutral squirmer ($\sim r^{-3}$). For a pusher or puller squirmer, a force dipole field is added ($\sim r^{-2}$), whereas $\chi\neq 0$ 
	refers to the rotlet dipole. Pushers model propulsion
	originating from the the back of a swimmer's body, such as 
	rotating flagella, whereas pullers are a more suitable description for a breast-stroke type of motion. Pusher and puller fields are the most typical hydrodynamic modes considered in microswimmer systems ~\cite{BerkeLauga2008,RuehleStark2020}. Furthermore, the 
	rotlet dipole field is an important aspect of many swimming organisms, for example, of bacteria with counter-rotating flagella and cell body. It has been studied in some previous works~\cite{FaddaYamamoto2020,IshimotoGaffney2013}.
	
	The complete hydrodynamic flow field generated and experienced by a collection of squirmers is strongly determined by
	their manifold hydrodynamic interactions and is therefore analytically untractable. However, we can still gain insights from 
	the single squirmer 
	fields and their different contributions, which can be ordered according to their radial decay. This can  
	help to interpret numerical results of their collective dynamics.
	
	In our case, the self-created flow fields of the squirmers from eq.~(\ref{eq:squirmer_field}) are not the only contributions.
	Also their gravitational force and bottom-heaviness generate flow fields. Furthermore, the no-slip 
	boundary condition on the bottom surface affects the hydrodynamics by creating a wall-induced flow
	\cite{SpagnolieLauga2012}. Fortunately, the Stokes equations are linear, which allows us to superimpose all occuring fields.
	
	\paragraph{Contributions from gravity}
	The first part comes from the gravitational force acting on the squirmer and is the same as the flow field of a passive
	sphere dragged through the fluid. Therefore, a squirmer under gravity always induces the flow fields of
	a stokeslet and a source dipole~\cite{KimKarrila2013}. The latter  adds to the source dipole of the squirmer \cite{PakLauga2014}. Additionally, since the center of mass of the squirmer is shifted by a distance $r_0$ from 
	its geometric center, it experiences bottom heaviness,
	which aligns the swimming direction $\mathrm{e}$ along the vertical $\mathbf{e}_z$. Due to the gravitational torque $\mathbf{T}_\mathrm{bh} = mgr_0 \mathbf{e} \times \mathbf{e}_z$ acting on the squirmer, 
	it rotates with angular velocity $\mathbf{\Omega}_\mathrm{bh} = \mathbf{T}_\mathrm{bh} / (8\pi \eta R^3)$ and thereby
	induces a rotlet field. The velocity fields of both contributions are~\cite{SpagnolieLauga2012,PakLauga2014,RuehleStark2020}:
	\begin{equation}
		\label{eq:gravity}
		\mathbf{u}_\mathrm{grav}(\mathbf{r}) = \frac{1}{4}v_\mathrm{sed} \left[ -3 \frac{R}{r}\left(\mathbf{e}_z + \frac{z}{r}\mathbf{\hat{r}}\right) + \frac{R^3}{r^3}\left(-\mathbf{e}_z + 3\frac{z}{r}\mathbf{\hat{r}}\right) \right]
	\end{equation}
	\begin{equation}
		\label{eq:bottom_heavy}
		\mathbf{u}_\mathrm{bh}(\mathbf{r}) = \frac{3}{4}v_0 \frac{r_0}{R \alpha}\frac{R^2}{r ^2}\left((\mathbf{e}\cdot\hat{\mathbf{r}})\mathbf{e}_z - \frac{z}{r}\mathbf{e}\right) \,. 
	\end{equation}
	Here, we introduced the dimensionless parameters 
	$\alpha = v_0 / v_\mathrm{sed}$ and $r_0/ (R\alpha)$,
	with the bulk sedimentation velocity $v_\mathrm{sed}=mg/(6\pi\eta R)$ and the center-of-mass shift $r_0$. 
	The parameter $\alpha$ measures the swimming velocity relative to the sedimentation velocity and 
	$r_0 / (R \alpha)$ is a unitless strength of the gravitational torque.
	In these units the related angular velocity becomes
	\cite{WolffStark2013}:
	\begin{equation}
		\label{eq:bh_angular}
		\mathbf{\Omega}_\mathrm{bh} = \frac{3}{4}\frac{v_0}{R}\frac{r_0}{R\alpha} \mathbf{e}\times \mathbf{e}_z \, .
	\end{equation} 
	The actual flow field of the squirmer together with the gravity-induced contributions constitute the hydrodynamic signature 
	of our problem and give the relevant flow fields in leading order in $1/r$.
	
	Importantly, the non-zero vorticities of these flow fields reorient nearby squirmers or the squirmer itself when it interacts 
	hydrodynamically with a wall. Since the vorticity of a source dipole is zero, we only need to consider the vorticity from the 
	gravity-induced stokeslet and rotlet
	as well as the vorticity from the force dipole of squirmers with $\beta\neq 0$. Using the definition of the vorticity, 
	$\boldsymbol{\omega}(\mathbf{r}) = \frac{1}{2}\nabla \times \mathbf{u}(\mathbf{r})$, the respective
	vorticities of the stokeslet, rotlet, and force dipole of a bottom-heavy squirmer are
	\begin{equation}
		\label{eq:vort_stokeslet}
		\boldsymbol{\omega}_{S} = \frac{3}{4}v_\mathrm{sed}\frac{R}{r^2}\mathbf{\hat{r}}\times\mathbf{e}_z,
	\end{equation}
	\begin{equation}
		\label{eq:vort_rotlet}
		\boldsymbol{\omega}_{R} = \frac{3}{4}v_\mathrm{0}\frac{r_0}{R\alpha}\frac{R^2}{r^3}\left(2\mathbf{e}\times\mathbf{e}_z - 3 \left((\mathbf{e}\cdot\mathbf{\hat{r}})\mathbf{\hat{r}}\times\mathbf{e}_z - \frac{z}{r}\mathbf{\hat{r}}\times\mathbf{e}\right)\right),
	\end{equation}
	\begin{equation}
		\label{eq:vort_fdp}
		\boldsymbol{\omega}_{Fd} = -v_0\beta\frac{R^2}{r^3}(\mathbf{e}\cdot\mathbf{\hat{r}})\left(\mathbf{e}\times \mathbf{\hat{r}}\right).
	\end{equation}
	Note that 
	the vorticity of the rotlet, $\boldsymbol{\omega}_{R}$ from eq.\ (\ref{eq:vort_rotlet}),
	vanishes for $\mathbf{e}\rightarrow\mathbf{e}_z$ and therefore this contribution is typically small.
	
	Now, placing a second squirmer at position $\mathbf{r}_2$ in the flow field of a sufficiently distant first squirmer, its
	angular and translational velocities are determined by the collected far field expressions.
	Using Fax\'{e}n's theorem \cite{KimKarrila2013,Faxen1922}, 
	the angular velocity  up to terms proportional to $1/r_2^3$
	is given by $\boldsymbol{\omega}_{S}(\mathbf{r}_2) + \boldsymbol{\omega}_R(\mathbf{r}_2) + 
	\boldsymbol{\omega}_{Fd}(\mathbf{r}_2) + \mathcal{O}(r_2^{-4})$ and the translational velocity
	becomes
	$\mathbf{u}_\mathrm{sq}(\mathbf{r}_2) 
	+ \mathbf{u}_\mathrm{grav}(\mathbf{r}_2)$
	$+ (R^2/6)$ $\nabla^2 \mathbf{u}_\mathrm{grav}(\mathbf{r}_2) + \mathbf{u}_\mathrm{bh}(\mathbf{r}_2) + \mathcal{O}(r_2^{-4})$, 
	where $\mathbf{r}_2$ is the distance vector from the first squirmer
	to the second.

	The collective dynamics in gyrotactic systems is 
	determined by the competition and balancing of external torque and flow vorticity acting on the microswimmers. 
	Assuming two squirmers on the same height with distance $r$, they rotate each other away from
	the vertical by an angle $\vartheta$ due to the stokeslet vorticity of eq.\ (\ref{eq:vort_stokeslet}), in leading order. Balancing this rotation 
	by the angular velocity of eq.\ (\ref{eq:bh_angular}) due to the gravitational torque gives
	\begin{equation}
		\frac{3}{4}\frac{v_0}{R}\frac{r_0}{R\alpha}\sin\vartheta = \frac{v_0}{\alpha}\frac{R}{r^2} \, .
		\label{eq.balance}
	\end{equation}
	In the following we will identify stable clusters of squirmers in our simulations only when the gravitational
	torque is sufficiently large. We present a simple argument for this behavior. At the closest possible distance $r=2R$,
	the angle where both angular velocities in eq.\ (\ref{eq.balance}) balance each other becomes
	\begin{equation}
		\sin\vartheta = \frac{1/(3\alpha)}{r_0/(R\alpha)} \, .
	\end{equation}
	Since $\sin \vartheta \le 1$, such a stable balance requires the lower bound
	\begin{equation}
		\label{eq:torque_condition}
		r_0/(R\alpha) \geq (3\alpha)^{-1}
	\end{equation}
	for the dimensionless torque.
	As we will show in Sect.\ \ref{sec:cluster_size} this gives a good estimate for the torque, where the clusters
	start to become compact.
	
	\paragraph{Wall terms}
	Close to a no-slip wall mirror multipoles have to be added to the hydrodynamic multipoles introduced above, in order to fulfill the no-slip boundary condition~\cite{SpagnolieLauga2012}. 
	The leading far-field contribution of a squirmer unter gravity is the stokeslet, which close to a no-slip wall
	turns into the Blake tensor~\cite{Blake1971Proceedings,BlakeChwang1974};
	the influence of its mirror multipoles decays with the inverse height. Furthermore, the squirmer type further affects the vertical velocity component of the wall-induced flow field, which depends on the squirmer orientation. For example, vertically oriented pushers are repelled and pullers are attracted to the wall~\cite{RuehleStark2018}.
	
	We are mainly interested in the vorticities 
	of the wall-induced flow fields since they are directly related to gyrotaxis. 
	We give the most relevant terms here following Refs.~\cite{RuehleStark2018,SpagnolieLauga2012}.
	By symmetry the mirror image in the Blake tensor does not contribute to the wall-induced vorticity acting on a single squirmer. Therefore, the leading-order contribution for pushers and pullers is the reflected force dipole field, whereas for $\beta = 0$ and $\chi \neq 0$
	it is the reflected rotlet-dipole field. 
	Second, a force quadrupole arises as a wall-induced mirror term in the three squirmer multipoles with strengths
	$B_1$, $B_2$ and $C_2$. 
	In particular, this force quadrupole results from the reflected vorticity field of the source dipole \cite{SpagnolieLauga2012} and, therefore, determines the orientation of the 
	neutral squirmer in wall proximity~\cite{RuehleStark2018}. We write the vorticity or angular velocity
	$\boldsymbol \Omega^{\mathrm{wall}}$  as a function of the distance to the no-slip wall, $\Delta z$, and the angle $\vartheta$ of the squirmer orientation $\mathbf{e}$ 
	with respect to the vertical~\cite{SpagnolieLauga2012}. Using cylindrical coordinates, we arrive at
	the following components
	\begin{eqnarray}
		\label{eq:rdp_rho}
		\Omega^\mathrm{wall}_{\rho} & =  &\frac{81}{32} \frac{v_0}{R}  \chi\sin\vartheta\cos\vartheta \frac{R^4}{\Delta z^4} \\
		\label{eq:rdp_phi}
		\Omega^\mathrm{wall}_{\phi} & = & -\frac{3}{16} \frac{v_0}{R} \sin\vartheta \left(\frac{R^4}{\Delta z^4} + 
		\frac{3}{2}\beta\cos\vartheta\frac{R^3}{\Delta z^3}\right)\\
		\label{eq:rdp_z}
		\Omega^\mathrm{wall}_{z} & = & -\frac{27}{64} \frac{v_0}{R} \chi (1-3\cos^2\vartheta)\frac{R^4}{\Delta z^4} \, .
	\end{eqnarray}  
	For example, this angular velocity 
	becomes zero for a vertically oriented neutral squirmer with $\beta,\chi=0$. 
	The additional force-dipole flow field of pushers and pullers change 
	this stable orientation~\cite{RuehleStark2018}. Furthermore, it is known that the rotlet dipole
	induces the wall component $\Omega^\mathrm{wall}_{z}$ of eq.\ (\ref{eq:rdp_z}) that is responsible for 
	circular swimming~\cite{IshimotoGaffney2013,LaugaStone2006}.
	
	\subsection{Hydrodynamic simulations}
	\subsubsection{Multi-particle collision dynamics}
	
	\label{sec:mpcd}
	We use the mesoscopic simulation technique of multi-par\-ti\-cle collision dynamics~\cite{MalevanetsKapral1999,MalevanetsKapral2000,GompperWinkler2009}, which is well established for hydrodynamic simulations of microswimmers, in particular, squirmers \cite{DowntonStark2009,TaoKapral2010,GoetzeGompper2010,ZoettlStark2014,BlaschkeStark2016,KuhrStark2017,TheersGompper2018,MandalMazza2021}.
	It simulates hydrodynamic flow fields as solutions of the Navier-Stokes equations and also includes thermal noise~
	\cite{PaddingLouis2006,GoetzeGompper2007}. The fluid
	consists of coarse-grained but point-like fluid particles of mass $m_0$. The algorithm of multi-particle collision dynamics applies a 
	sequence of alternating streaming and collision steps to these particles~\cite{GompperWinkler2009}. 
	
	During the streaming step with 
	duration $\Delta t$, the fluid particles move ballistically. During this step one also implements the no-slip and slip boundary conditions at the respective bounding walls and squirmer surfaces using the bounce-back rule~\cite{PaddingLouis2005}. Furthermore, the 
	positions of the suspended squirmers are updated using 
	velocity-Verlet integration~\cite{ZoettlStark2018}, 
	which accounts for the gravitational forces and torques acting on them. For the collision step we introduce a length scale $a_0$ to indicate the 
	range of fluid-particle collisions, which take place in cubic cells of side length $a_0$. On average, we place $n_\mathrm{fl}$ fluid 
	particles into the cells. Both values of $\Delta t$, as well as $n_\mathrm{fl}$, determine the fluid properties such as viscosity. 
	Further simulation details are described more exhaustively in our previous works~\cite{RuehleStark2020,BlaschkeStark2016,ZoettlStark2018}.
	
	The collisions of the fluid particles inside a collision cell are governed by a specific collision operator or collision rule. Several different collision rules exist for multi-particle collision dynamics~\cite{GompperWinkler2009}. We use the MPC-AT+a algorithm which assures Galilean invariance and angular momentum conservation~\cite{GompperWinkler2009,NoguchiGompper2007}. 
	It is based on the Andersen thermostat~\cite{GompperWinkler2009,GoetzeGompper2007}, which 
	keeps the thermal energy of the fluid at the fixed value
	$k_B T_0$. Thus, the cell length $a_0$, the mass $m_0$ of a fluid particle, and the thermal energy $k_BT_0$ form the units of
	length, mass, and energy. From this we can derive other units, in particular, 
	for time $\sqrt{m_0/k_BT_0}$ and velocity $\sqrt{k_BT_0/m_0}$.
	
	We choose the Reynolds number as small as computationally feasible
	and arrive at $\mathrm{Re}=0.17$~\cite{RuehleStark2020}. This is still a good approximation of  the Stokes flow regime.

	\subsubsection{Implementation and parameters}
	We simulate the MPCD fluid inside a rectangular system with height $H$ and a quadratic cross section with edge length $L$. We use periodic boundary conditions in lateral direction and two no-slip walls in vertical direction. Usually, we have $H=L=108a_0$, \textit{i.e.}, 
	a cubic simulation box.
	For the fluid we use the duration of the streaming step 
	$\Delta t = 0.02\sqrt{m_0/k_BT_0}$ and the fluid-particle number density $n_\mathrm{fl}=10$. This choice of parameters results in a viscosity of $\eta=16.05\sqrt{m_0k_BT_0}/a_0^2$~\cite{ZoettlPHD}. Our squirmers have $B_1=0.1\sqrt{k_BT_0/m_0}$ and, unless 
	stated otherwise, $\beta=\chi=0$. The resulting swimming velocity is $v_0=0.067\sqrt{k_BT_0/m_0}$, which corresponds to
	an active P\'{e}clet number $\mathrm{Pe}=\frac{Rv_0}{D_T} = 330$, comparable, for example, to bacteria~\cite{JanosiHorvath1998,ChattopadhyayWu2006}. Furthermore, the squirmers have a radius $R=4a_0$, and the relevant
	ballistic time scale $v_0/R$ amounts to around $3000\Delta t$~\cite{RuehleStark2020}.
	To set  the parameter $\alpha$, we choose a moderate strength of gravity that always creates a sedimentation velocity larger than $v_0$, typically $\alpha=0.8$. This prevents squirmers from escaping from the bottom of the system 
	to the top wall.
	
	Due to the high computational cost of hydrodynamic simulations, we compile software for running on multiple processing units simultaneously, \textit{i.e.}, the program is significantly parallelized. Earlier simulations were performed on a high-perfomance computing cluster using 
	approximately 150 CPU cores. For later simulations we switched to a CUDA implementation, which we ran on graphics processor units.
	
	\paragraph{Initial condition}
	We tested two different initial conditions. Either, the squirmers were distributed randomly in the simulation box or we initialized 
	them on a regular grid in the midplane $z=H/2$, all pointing upwards
	with $\cos\vartheta=1$. This had little effect on the clustering behaviour that we focus on in the following. The only difference we observed was 
	that for the random initialization more squirmers 
	escaped to the top wall for strong bottom-heaviness. 
	Otherwise, these squirmers have no influence on the rest of the simulation, as they stay 
	at the top plate without interfering during the rest of the simulation. Choosing random orientations at the beginning of the simulation or initializing the squirmers in another plane with constant $z$ also brought no qualitative change.

	\section{Results}
	\label{sec:results}
	
	We present our results on gyrotactic clusters and the associated parameter studies in the following. The decisive parameters are the 
	velocity ratio $\alpha$ and the 
	torque measure $r_0/R\alpha$ that we introduced above. In previous studies~\cite{RuehleStark2020,KuhrStark2017} we investigated large-scale convection and plumes that move over the entire system height. In contrast, we here look at the cluster formation of bottom-heavy squirmers that are constrained to the bottom
	wall due to strong gravity ($\alpha <1$). However, the squirmers are still capable of moving vertically, in contrast to monolayer formation for $\alpha$ below $0.1$~\cite{KuhrStark2019}. Thus, we observe the emergence of dynamic clusters from a squirmer suspension that float above the bottom wall.
	
	\subsection{Phenomenology of gyrotactic clusters}
	
	At the outset we discuss the basic properties of the emergent structures that form under gravity and bottom-hea\-vi\-ness.
	We present snapshots of the squirmer system and discuss the 
	vertical density profiles and dynamics of the clusters.
	However, we start by illustrating the coupled dynamics of two squirmers.

	\subsubsection{Gyrotaxis of a squirmer pair}
	\label{subsubsect.pair}
	
	\begin{figure}
		\centering
		\resizebox{0.45\textwidth}{!}{\includegraphics{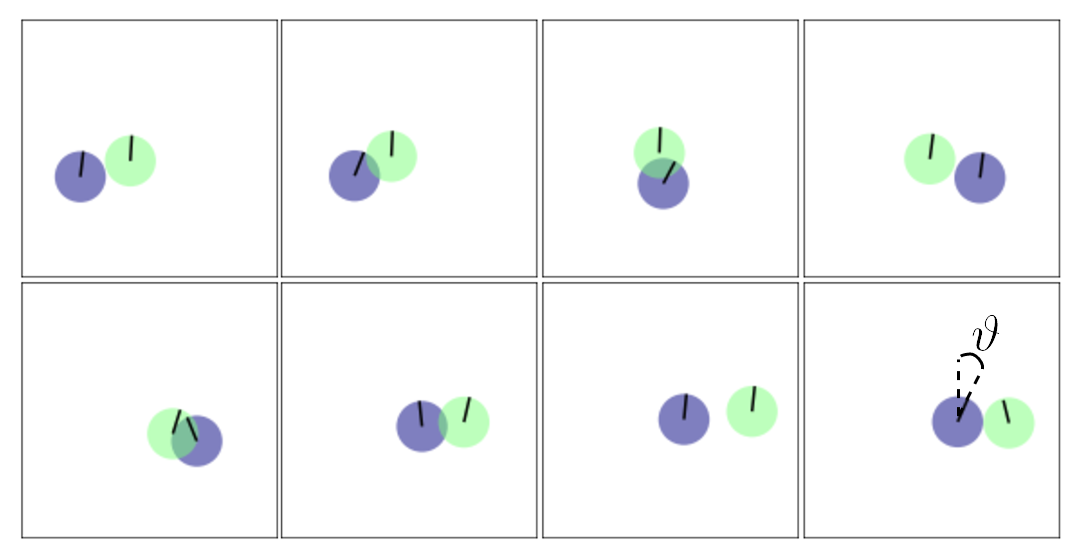}}
		\caption{Snapshots of a pair 
			of squirmers performing coupled oscillations above the bottom wall. The simulations used the 
			parameters $\alpha=0.8$ and $r_0/R\alpha=0.16$.
		}
		\label{fig:pair_snap}
	\end{figure}
	
	Gyrotaxis is often discussed in the context of collective dynamics of microswimmers, but already a two-particle system can illustrate the combined effect of a gravitational torque and  fluid flow.
	For example, Drescher \emph{et al.} calculated the 
	coupled trajectories of two
	microswimmers based on eqs.\ (\ref{eq:gravity}), (\ref{eq:bh_angular}), and (\ref{eq:vort_stokeslet}), just by considering the stokeslet flow fields and the 
	gravitational torque, and thereby reached good agreement with 
	their experimental data~\cite{DrescherGoldstein2009}.
	They found that the dynamical system of two 
	hydrodynamically interacting microswimmers contains oscillatory
	trajectories when reorientation due to the
	neighbor's stokeslet vorticity and due to the own gravitational torque balance each other. In the following, we reproduce such oscillations in our squirmer simulations.
	We show the simulated oscillations for a pair of squirmers in Fig.~\ref{fig:pair_snap} and outline the mechanism in the following.

	The simulations are performed for $\alpha=0.8$, where single neutral squirmers float above the bottom wall at a height of several squirmer radii $R$~\cite{RuehleStark2018}. 
	Due to bottom-heaviness, they experience a gravitational
	torque proportional to $-\sin\vartheta$, where
	$\vartheta$ is the polar angle between squir\-mer orientation and vertical
	(see last snapshot in the bottom row 
	of Fig.\ \ref{fig:pair_snap}).
	Thus, squirmers in bulk tend towards the vertical where the 
	gravitational torque is zero. Two squirmers with such a vertical orientation are shown in the first snapshot in Fig.~\ref{fig:pair_snap}. 
	The neighbor can tilt the orientation vector towards itself
	due to the vorticity of the stokeslet, as shown in the second snapshot. Balancing reorientations due to vorticity and torque, the squirmers are tilted and thus move towards each other. They eventually 
	pass each other (snapshots 3 and 4), whereupon the sense of the vorticity reverses and the motion repeats itself in the other direction, which is shown in snapshots 5-8. The external torque is important to stabilize the pair so that the distance of the squirmers does not increase too much after they have passed each other, which would 
	otherwise destroy the oscillations \cite{DrescherGoldstein2009}.  
	Interestingly, even for large torques the 
	squirmer orientations are not completely frozen, but pair formation  still occurs with a smaller oscillation amplitude. Since the gravitational torque approaches zero for a vertical orientation, 
	a small tilt and thereby horizontal motion can still be achieved by the vorticity in the flow field. 
	
	\subsubsection{Formation of clusters}
	
	\begin{figure}
		\centering
		\resizebox{0.49\textwidth}{!}{\includegraphics{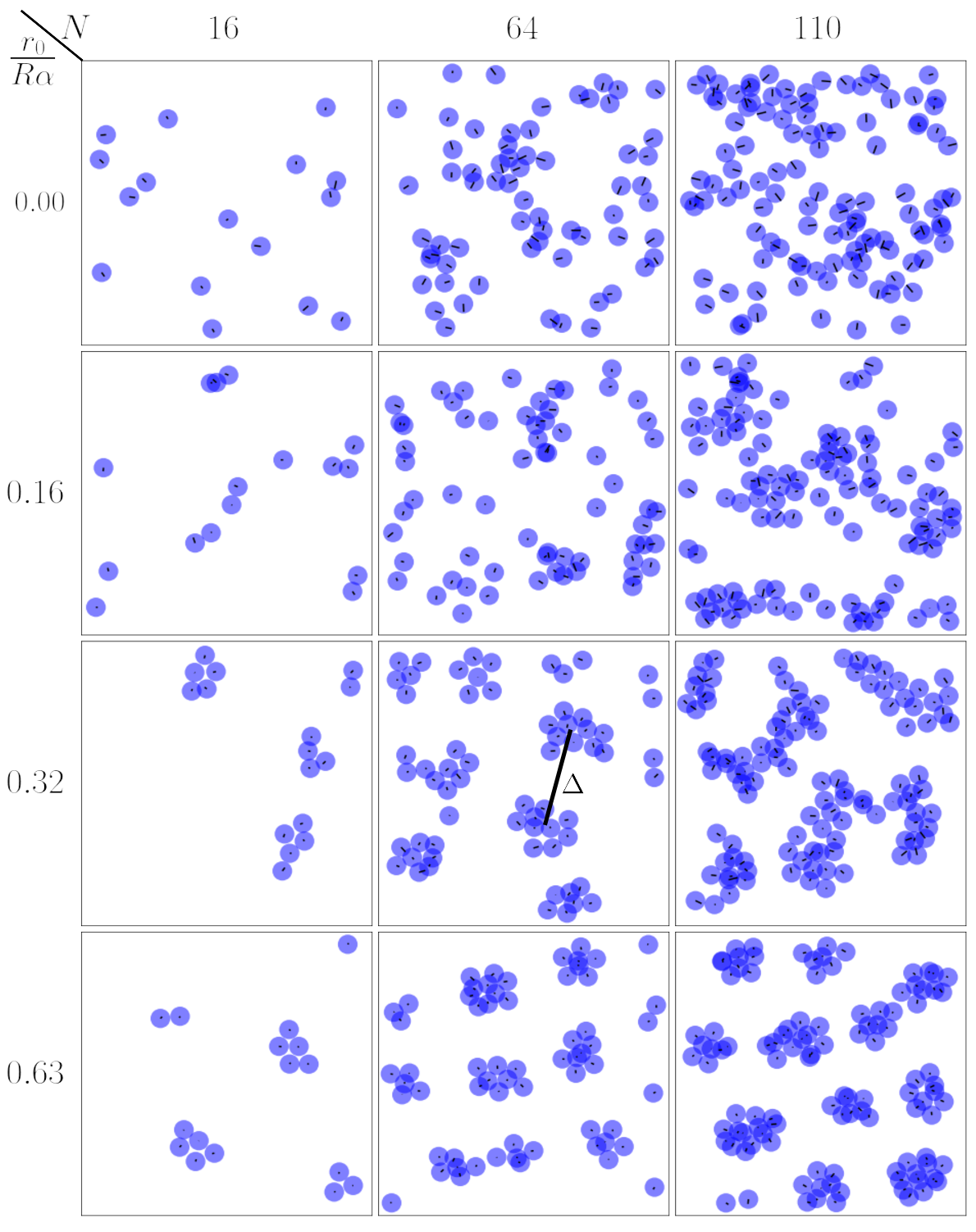}}
		\caption{Top view of snapshots of
			squirmer configurations. They show clusters that form in
			simulations with different 
			numbers of squirmers $N$ and torque strengths $r_0 / (R\alpha)$. Other parameters are $\beta = 0$ and $\alpha=0.8$.
		}
		\label{fig:cluster_snap}
	\end{figure}
	
	Now, we give an overview over what happens to the oscillatory motion in systems with a larger number of neutral squirmers.
	In Fig.~\ref{fig:cluster_snap} we show 
	snapshots
	of squirmer configurations seen from the top for different 
	torque strengths and squirmer numbers. We used a constant $\alpha=0.8$. The third row with $r_0/R\alpha=0.32$ is available in the supplement as videos M1-M3. These videos also show a side view of the system.
	
	The snapshots show that squirmer clusters emerge 
	for different squirmer numbers and torque values and spread over the whole plane with a typical cluster distance $\Delta$
	as indicated in the central snapshot for $r_0/R\alpha=0.32$. 
	These clusters only emerge at finite
	gravitational  torques. At zero torque (top row), squirmers 
	briefly touch during collisions but do not form
	clusters.
	Video M4 shows this case for $N=64$. Without the 
	gravitational torque and the strong tendency to align with the vertical, only hydrodynamic interactions and thermal noise determine the orientations of squirmers. 
	They tilt more strongly and, therefore, exhibit
	stronger horizontal motion.
	As explained above, squirmers swim past each other and cannot form stable structures.
	
	\begin{figure}
		\centering
		\resizebox{0.45\textwidth}{!}{\includegraphics{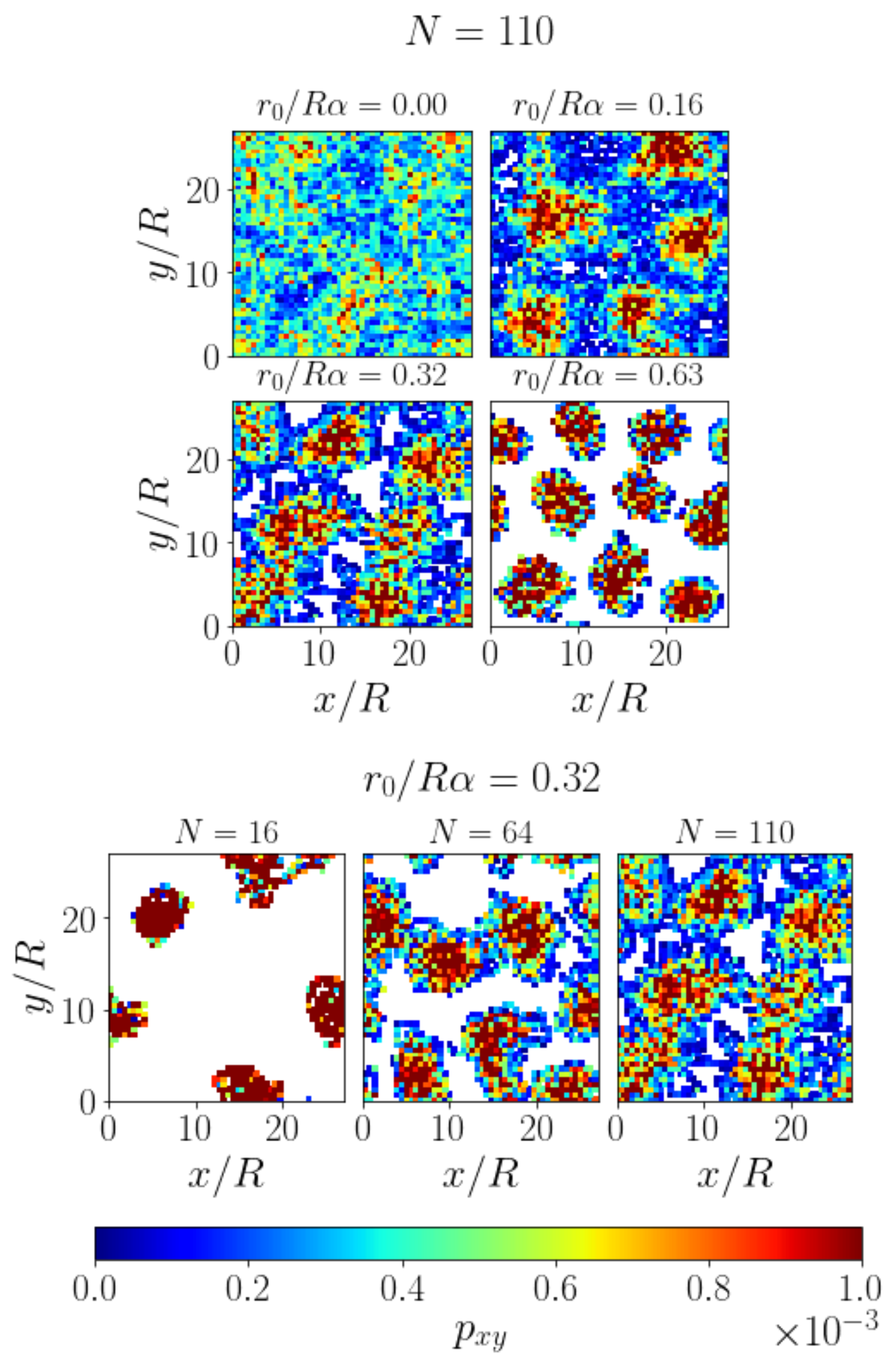}}
		
		\caption{
			Density profiles after averaging	
			over $10^5$ time steps. 
			Top: for different torque values and $N=64$ squirmers.
			Bottom: for different $N$ (area fractions) and torque value $r_0/R\alpha=0.32$.
		}
		\label{fig:2dhist}
	\end{figure}
	
	In Fig.~\ref{fig:2dhist} we show
	the density profiles of the squirmer system in the horizontal plane. In the top row the torque increases from left
	to right while $N=64$.
	The density profiles were obtained by averaging over $10^5$ time steps and over all squirmers in the bottom half of the system.
	At zero torque the 
	density profile is unstructured and almost uniform. There is some patchiness which we attribute to the aligning effect of the bottom wall, leading to floating squirmers that stay in the same spot for longer times~\cite{RuehleStark2018}.
	
	At $r_0/R\alpha=0.16$ we observe 
	that squirmers form clusters which are stable during a longer time
	(Fig.\ \ref{fig:2dhist}, top).
	There is still considerable horizontal motion and 
	squirmers frequently join and leave the clusters,
	therefore the space around the density peaks is still well explored.
	With increasing torque (\textit{e.g.}, 
	last two rows in Fig.~\ref{fig:cluster_snap}) the clusters become more static and compact. This is understandable since a stronger vertical alignment obstructs the escape of squirmers from clusters because 
	the horizontal velocities are smaller. According to the two-dimensional density profile for $r_0/R\alpha=0.32$
	in Fig.\ \ref{fig:2dhist}, top, the squirmers explore less space in the horizontal plane and the clusters indeed become more compact, 
	which is most clearly seen for $r_0/R\alpha=0.63$.
	The same effect occurs for the oscillating squirmer pairs, \textit{i.e.}, 
	the horizontal extent of the clusters decreases with 
	increasing torque.
	In parallel,  the total number of clusters increases as the right-most column of Fig.~\ref{fig:cluster_snap} shows for $N=110$.

	The bottom plot of Fig~\ref{fig:2dhist} shows how 
	the clustering of squirmers evolves when 
	their number or area fraction increases
	at constant torque.
	We observe a stronger horizontal motion, 
	which is visible from the disappearance of unvisited space and 
	less pronounced peaks in density. This makes sense since each additional squirmer introduces a new stokeslet, 
	which reorients neighbors away from the vertical.
	
	\paragraph{Cluster dynamics}
	Drescher \emph{et al.} coined the term ``minuet dance" for the
	oscillating two-squirmer motion
	\cite{DrescherGoldstein2009}.
	For larger clusters the motion 
	looks less regular and consists of squirmers switching their positions inside the cluster as videos M1-M3 show. We stress again that the clusters become more static with increasing 
	acting torque.
	The cluster reorganization sometimes creates
	symmetrical structures, such as pentagons or hexagons, as 
	video M2 illustrates. 
	However, these configurations do not 
	persist.
	A similar effect was reported in experiments with active emulsion droplets \cite{HokmabadPHD}.
	They form clusters that also rearrange frequently at high P{\'e}clet number.
	
	\subsubsection{Vertical density profile}
	
	\begin{figure} 
		\centering
		\resizebox{0.4\textwidth}{!}{\includegraphics{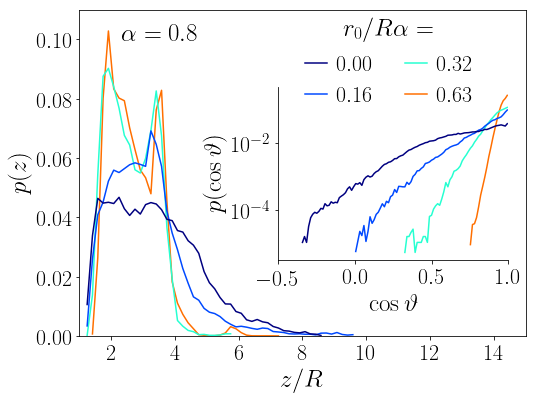}}
		
		\caption{
			Vertical density profile plotted versus squirmer height $z/R$ for different torque values at $\alpha = 0.8$
			and for $N=64$ squirmers. Inset: Distribution functions of vertical orientation angle $\theta$.
		}
		\label{fig:height_ori_hist}
	\end{figure}
	
	\paragraph{Stacking}
	The top-view snapshots in Fig.~\ref{fig:cluster_snap} show occasional stacking of squirmers, visible as overlaps, especially at the higher squirmer number in the right column. In order to investigate the vertical structure more closely, we show the vertical density profiles in Fig.~\ref{fig:height_ori_hist} for different torque values.
	The stacking can clearly be seen as a double peak at 
	the two higher torque values.
	More precisely, a bilayer is formed since the distance between the two peaks in Fig.~\ref{fig:height_ori_hist} for $r_0/R\alpha=0.32$ 
	and $0.63$ is approximately one squirmer diameter.
	
	\paragraph{Collective sinking}
	Note that the vertical density profile at zero rescaled torque
	reaches to larger heights compared to the non-zero values. Thus, the squirmers are more often found at larger heights.
	This is the case, even though the orientational distribution gets more strongly peaked at the vertical 
	orientation ($\cos\theta = 1$) with increasing $r_0/R\alpha$ (see inset of Fig.~\ref{fig:height_ori_hist}). The reason for this counter-intuitive behaviour is a reduced 
	translational friction coefficient per squirmer when they form clusters \cite{CichockiHinsen1995,ReichertStark2004}. 
	At zero torque clusters do not exist and squirmers sink individually due to gravity. In contrast, for higher torques
	they sediment faster within clusters
	since hydrodynamic friction is reduced due to hydrodynamic interactions,
	as was studied in detail in Ref.~\cite{RuehleStark2020}. This leads to stronger effective sinking, even though the alignment is more vertical.


	\subsection{Parameter study}
	\label{sec:parameters}
	
	We present a more detailed parameter study of squirmer clustering due to gyrotaxis. We characterize clusters by their radius 
	and mean particle number, look at the radial distribution function, study the influence of the squirmer flow field beyond the
	neutral squirmer, and look closer at the influence of gravity by varying $\alpha$.
	
	\subsubsection{Cluster radius and size}
	\label{sec:cluster_size}
	
	\begin{figure}[htp]
		\centering
		\resizebox{0.45\textwidth}{!}{\includegraphics{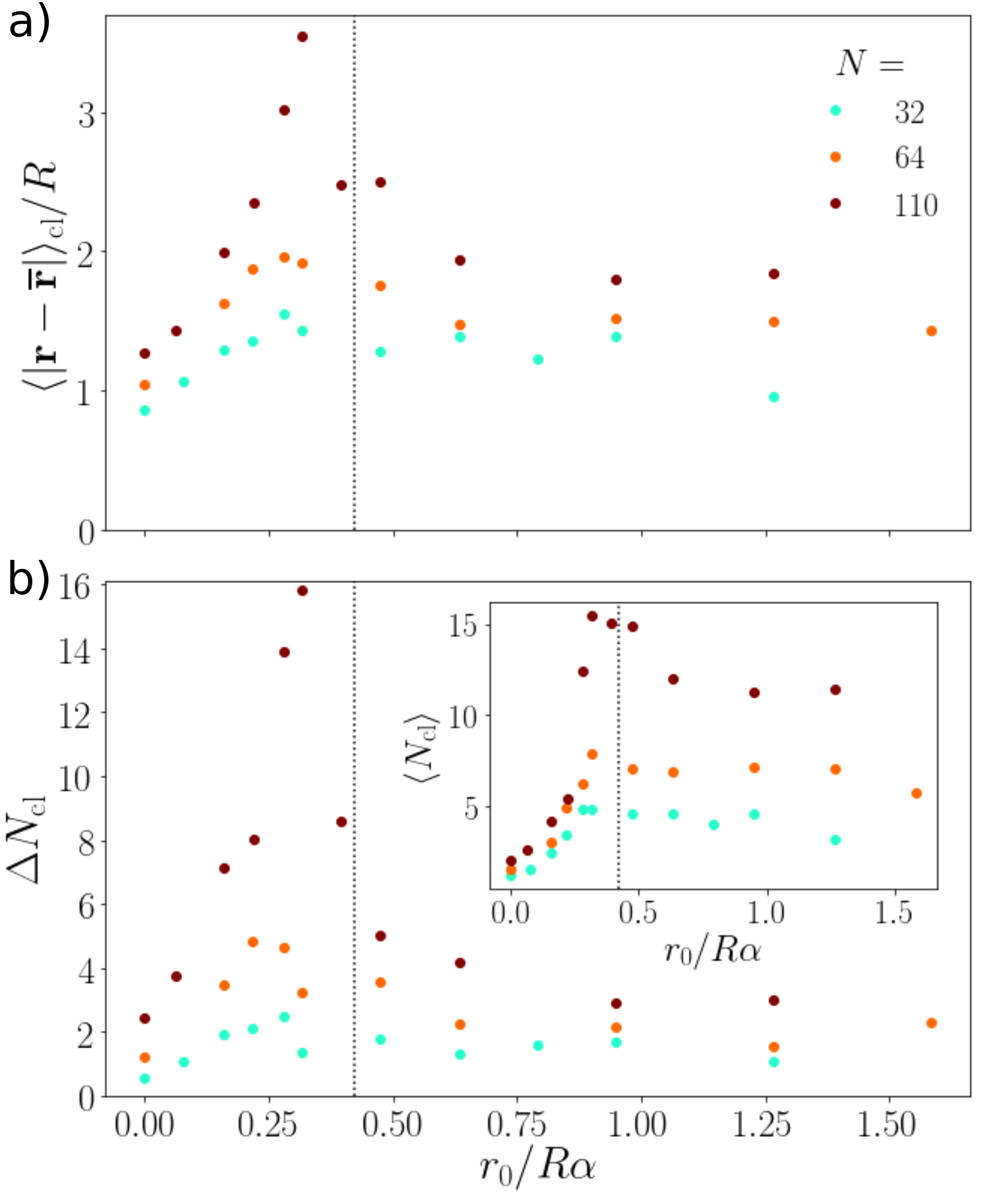}}
\caption{
	(a) Mean cluster radius $\langle \vert\mathbf{r}-\mathbf{\overline{r}}\vert\rangle_\mathrm{cl}$ in units of $R$
	plotted versus torque value $r_0/R\alpha$ for different squirmer numbers $N$. (b) Standard deviation 
	$\Delta N_\mathrm{cl}$ 
	and inset: mean number of squirmers in a cluster $\langle N_\mathrm{cl} \rangle$. The dotted vertical lines show      the equality condition of eq.~(\ref{eq:torque_condition}) for $\alpha=0.8$.
}
\label{fig:mean_cluster_size_torque}
\end{figure}

In order to 
demonstrate how the gravitational torque influences the sizes of the emerging clusters, we 
define measures for the cluster extension and size. We plot them versus the torque value in Fig.~\ref{fig:mean_cluster_size_torque}
for different squirmer numbers $N$. First, to quantify the cluster extension,
we introduce a mean cluster radius 
by determining the average distance of a squirmer 
to the cluster center at $\mathbf{\overline{r}}$,
$\langle \vert\mathbf{r}-\mathbf{\overline{r}}\vert\rangle_\mathrm{cl}$. Here, besides the time average, $\langle \ldots \rangle_\mathrm{cl}$ 
also means averaging over all squirmers in a cluster and then over  all clusters.
Second, for the cluster size we 
calculate the mean number of squirmers in a cluster,
$\langle N_\mathrm{cl} \rangle$.
We also consider the standard deviation $\Delta N_\mathrm{cl} = [\langle  (N_\mathrm{cl} - \langle N_\mathrm{cl} \rangle )^2 \rangle ]^{1/2}$
to have a measure how much the cluster size varies between the different clusters and also in time.
To determine all these quantities, we define a cluster
as a set of squirmers, where for each squirmer the 
distance to at least one other squirmer in this set is less than $2R +R/4$. Finally, we only take the last $2.5\cdot 10^5$ timesteps into account where the system has reached steady state.

Both, the cluster extension (a) and 
size [inset of (b)], show a pronounced maximum, 
especially for $N=110$, where clusters are largest. The increase at small torque values leading up to this maximum is quite sharp.
As already stated, in this regime clusters start to form due to gyrotaxis and are more short-lived,
because the external torque is not strong enough to stabilize them. 
At and close to the maxima in extension and size, the clusters are rather loosely bound,
which is visible in the snapshots for torque value $r_0/R\alpha = 0.32$ in Fig.\ \ref{fig:cluster_snap}
and in the corresponding videos M1-M3. The clusters strongly vary in size and also in shape with ongoing time.
This behavior is reflected by the peak in the variance $\Delta N_\mathrm{cl}$ of the mean cluster size 
as shown in the main plot of Fig.\ \ref{fig:mean_cluster_size_torque}(b).

Beyond the maxima of cluster radius and size, both quantities decrease with increasing torque. The clusters become more compact and exhibit less variation also in time, as the decrease in the standard deviation $\Delta N_\mathrm{cl}$ shows. This behavior is most 
clearly visible for $N=110$. For further increasing torque all three quantities then become nearly constant. While the decrease in $\langle N_\mathrm{cl} \rangle$ is not very pronounced, it is stronger for the mean cluster radius, in particular, for $N=110$. 
Thus the clusters become compactified, while their 
total number increases, as the last row in Fig.\ \ref{fig:cluster_snap} nicely
demonstrates. Interestingly, the compactification of the clusters agrees with the location of the dashed line. According to 
eq.\ (\ref{eq:torque_condition}) it shows the value of the torque that is just able to balance squirmer rotation due to the stokeslet 
vorticity of a neighboring touching squirmer. Since squirmers in a cluster also hinder each other from moving, the cluster is stabilized 
and forms compact objects.

The reason for the decrease of 
the cluster radius 
is similar to the two-squirmer case described in Sect.\ \ref{subsubsect.pair}. As the torque increases, squirmers are more vertically oriented and thus their horizontal mobility is further reduced. This has 
the effect that the motion of squirmers within each cluster is more constrained just as the oscillation amplitude of the squirmer 
pair also decreases  
with the torque. In order to follow the vorticity acting on the squirmers, they eventually also evade in
the negative $z$-direction and the cluster assumes a stacked arrangement, as we 
have seen in Fig.~\ref{fig:height_ori_hist}. 

\subsubsection{Radial distribution function}

\begin{figure}
\centering
\resizebox{0.4\textwidth}{!}{\includegraphics{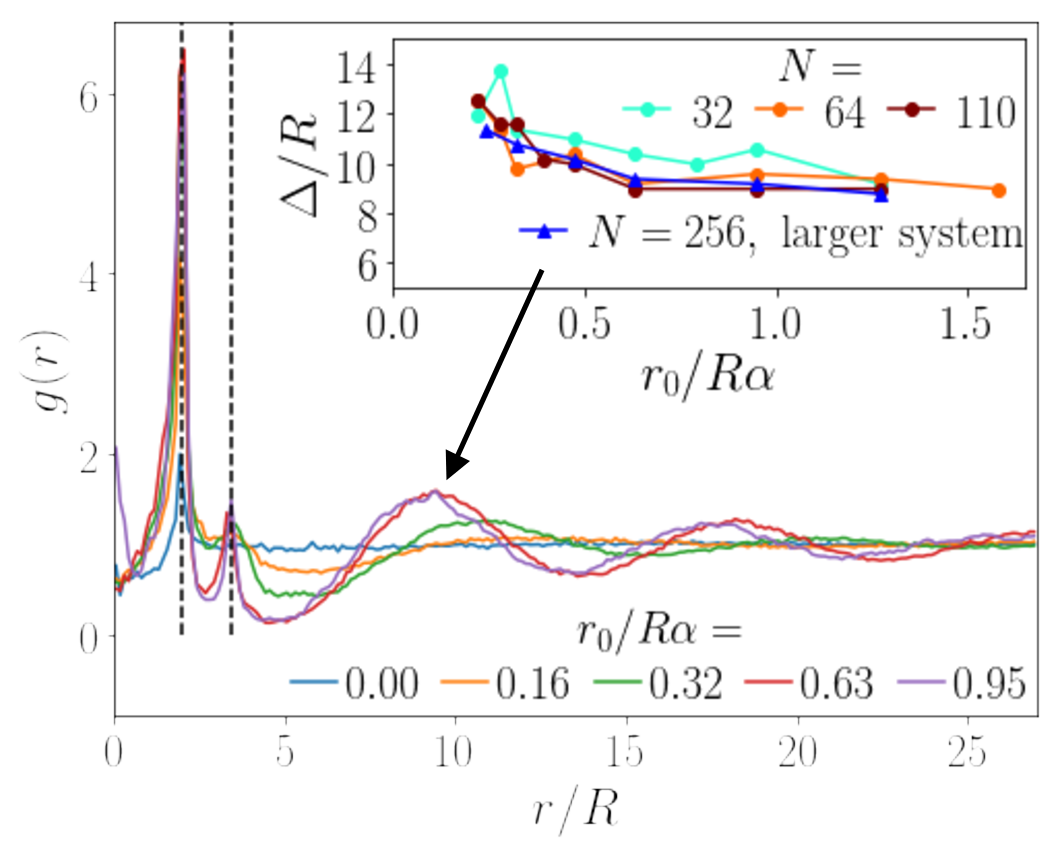}}
\caption{ Radial distribution function 
	for several non-dimensional torques for a larger system with $256$ squirmers and linear system size $L=216$.
	Inset: Mean cluster distance $\Delta$ 
	plotted versus torque for three squirmer numbers at the regular siystem size and $N=256$ at the larger system size. 
	The dashed lines indicate the nearest-neighbor and next-nearest-neighbor distances within a cluster
	at $2R$ and $2\sqrt{3}R$, respectively.}
\label{fig:rdf_maxima}
\end{figure}

To investigate squirmer ordering within the clusters and also how the clusters position themselves relative to each other, 
we calculate the radial distribution function $g(r)$. Here, $r$ is the squirmer distance in a 
horizontal plane, after projecting all squirmer positions on this plane. Thus, we treat the squirmer system as a two-dimensional system.
The two-dimensional radial distribution function for $N$ squirmers in an area $A$ with positions $\mathbf{r}_i$ is defined as~\cite{ChaikinLubensky1995}
\begin{equation}
g(r) = \frac{1}{N/A}\langle \sum_{i\neq j} \delta(\vert\mathbf{r}_i - \mathbf{r}_j\vert) \rangle,
\end{equation}
where $ r = | \mathbf{r}_i - \mathbf{r}_j | $ and the ensemble average is performed over all recorded squirmer configurations
in the last $2.5\cdot 10^5$ time steps of the simulations. Note that a constant value of $g(r)=1$ corresponds to the random ordering of an ideal gas.

The radial distribution function gives meaningful results for distances up to ca. $L/2$, thus half the linear box size.
Figure\ \ref{fig:cluster_snap} shows that the cluster distance $\Delta$ can assume such values. In order to rule out finite-size effects, 
we have doubled the linear size of our simulation box and use $L=216a_0$. This increases the cross-sectional area by a factor of $4$.
Since we simulate $256$ squirmers, the area fraction in the horizontal plane is the same as when we use $N=64$ squirmers for our regular box size.

In the main plot of Fig.~\ref{fig:rdf_maxima} we show the radial distribution function $g(r)$ for the larger system for several torque 
values. In all curves a peak at the nearest neighbor distance $r=2R$ is visible indicated by the dashed vertical line on the left. The blue 
curve for zero torque then assumes the constant value one at larger distances reflecting the randomly distributed squirmer positions. 
Already at $r_0/R\alpha=0.16$ a small maximum develops around the next-nearest neighbor distance $2 \sqrt{3}R=3.46 R$ 
(right dashed line), which becomes more pronounced with increasing torque. In particular, for the two highest torque values, where 
compact squirmer clusters occur, sharp peaks are visible. Note that both peaks are broadened towards smaller distances due to the 
partial overlap of the squirmers, when they assume different heights. The overlap is clearly visible in the snapshots of Fig.~\ref{fig:cluster_snap} as already discussed earlier.

Now we discuss the characteristic distances $\Delta$ between the clusters as revealed by the radial distribution function.
Already at $r_0/R\alpha= 0.16$ a weak minimum appears around $r=6R$ followed by a very shallow maxium located between $r=10R$ 
and $r=15R$. This signals the appearance of squirmer clusters separated from each other. Increasing the torque to 
$r_0/R\alpha=0.32$ (green curve), this third peak becomes more pronounced and its position shifts to smaller values. Finally, 
for the two highest torque values ($r_0/R\alpha = 0.63$ and 0.95), where clusters show compact packing, the peaks are identical. 
They are further shifted to smaller distances and their heights are larger. Interestingly, a further maximum for both torque values is 
visible at around $r=17R$ and $18R$, respectively. It corresponds to the next-nearest neighbor shell of the cluster ordering.
The maximum cannot be seen in the system with regular size, since all clusters are nearest neighbors due to the periodic boundary 
conditions.

Now, we focus on the the average cluster-cluster distance $\Delta$ in the inset of Fig.~\ref{fig:rdf_maxima}. We plot it versus 
the dimensionless torque for different squirmer numbers $N$  in our regular system size and for the larger system. For $\Delta$ we 
use the position of the first maximum in $g(r)$ connected to cluster ordering. With increasing $N$ the curves slightly shift downwards 
but overall we can state that the cluster distance does not depend on the area density of the squirmers. As we already realized, at small torques, when the clusters start to form, the cluster distance decreases with increasing torque but once compact clusters are 
formed at $r_0/R\alpha=0.64$ and beyond it stays constant.

\subsubsection{Influence of squirmer flow fields}

\begin{figure}[tp]
\resizebox{0.49\textwidth}{!}{\includegraphics{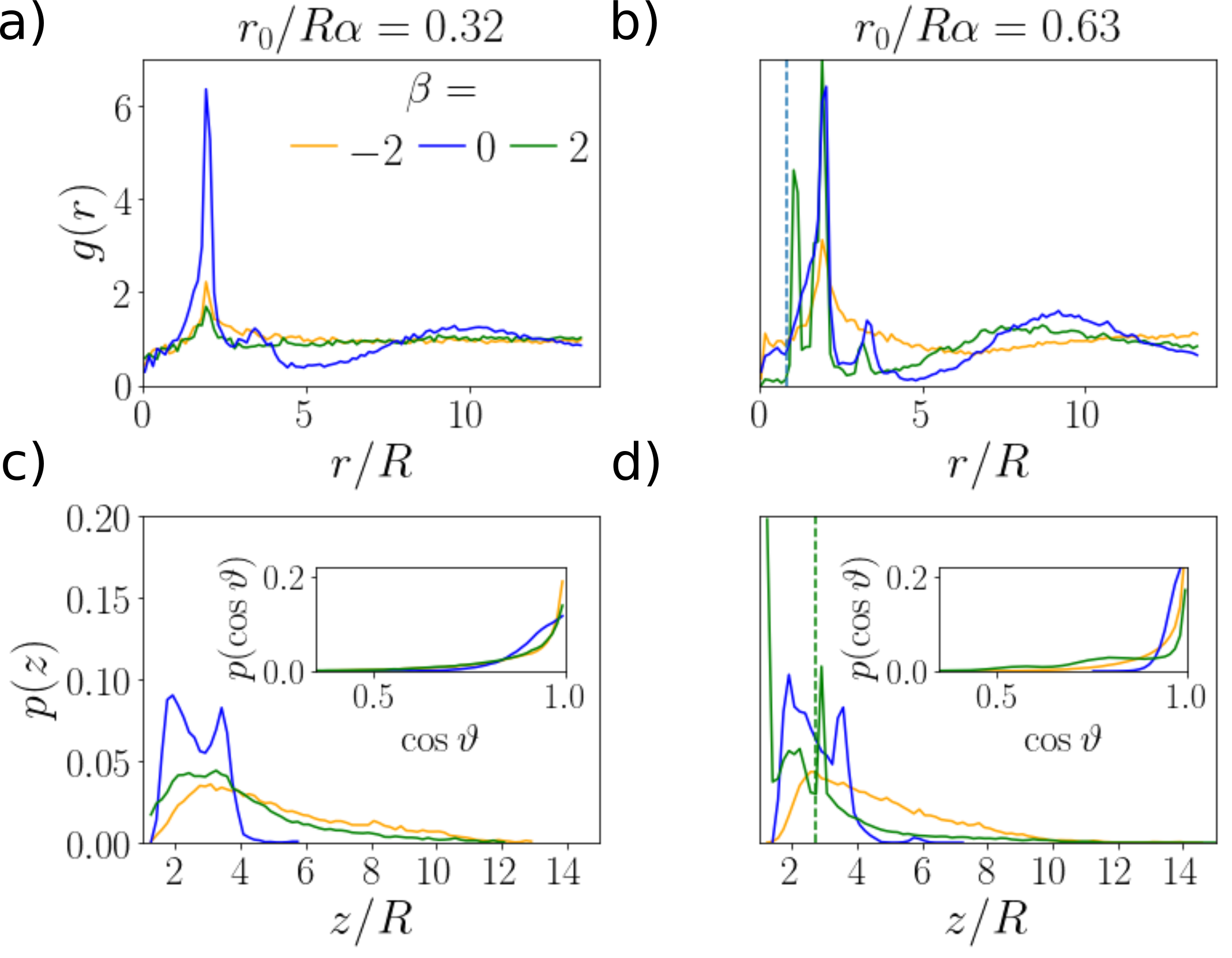}}
\caption{Radial distribution function for $\beta=-2,0,2$ at torque values a)
	$r_0/R\alpha= 0.32$ and b) $r_0/R\alpha= 0.63$.
	The regular system size with $64$ squirmers
	and $\alpha = 0.8$ is used. Vertical density
	profiles for the same parameters at c) $r_0/R\alpha= 0.32$ and d) $r_0/R\alpha= 0.63$.
	Insets: Distributions of vertical orientations.}
\label{fig:beta_hist}
\end{figure}

We consider two squirmer modes in addition to the $B_1$ mode that modify the squirmer type. First, the $B_2$ mode quantified 
by the parameter $\beta= B_2/B_1$ 
creates a force dipole in the far field and second, the $C_2$ mode introduces a rotlet dipole in the far field and is tuned via the parameter 
$\chi= C_2/B_1$, as we explain in Sect.~\ref{sec:flowfields}.

\begin{figure}
\resizebox{0.49\textwidth}{!}{\includegraphics{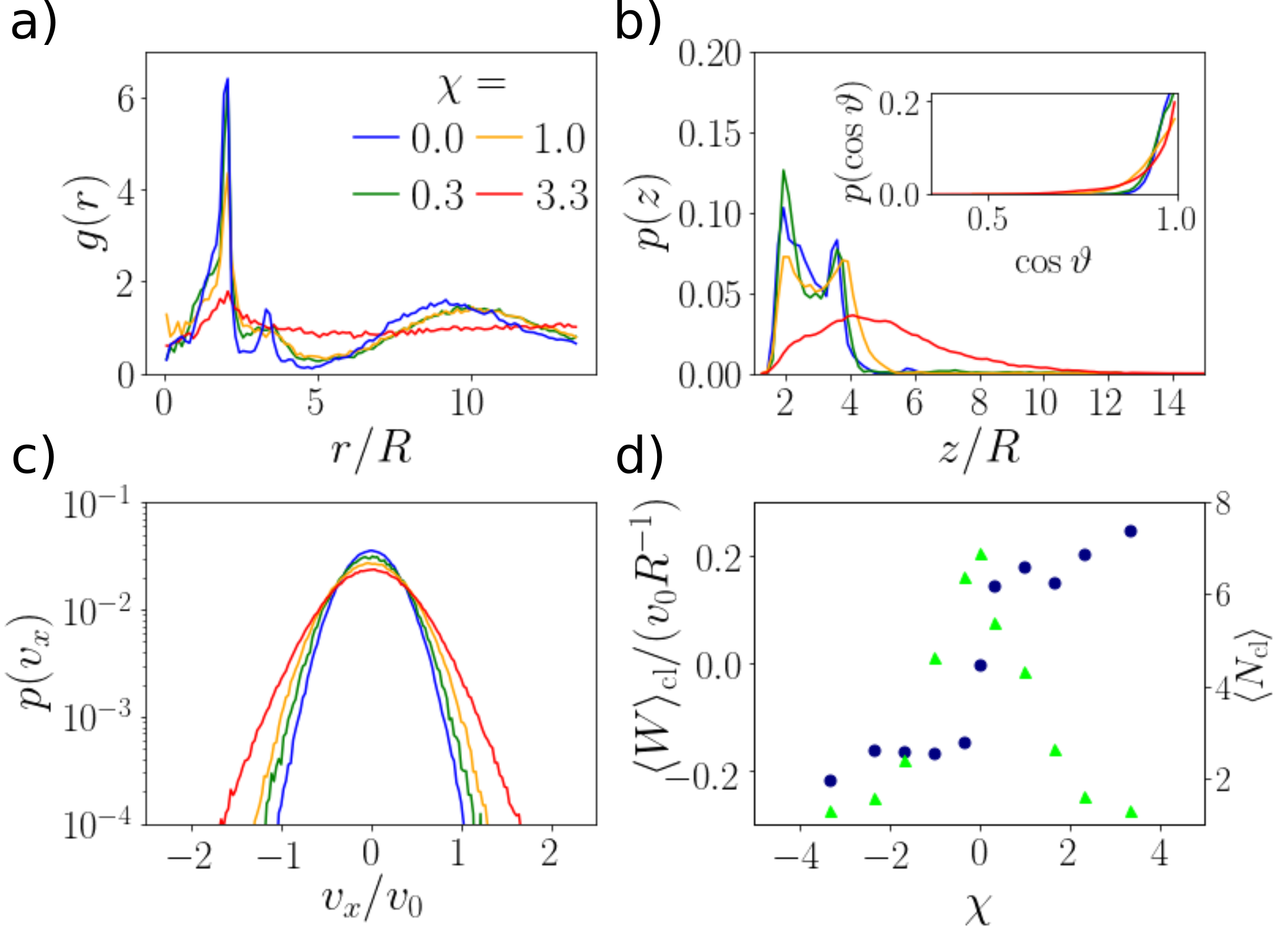}}
\caption{
	(a) Radial distribution function,
	(b) vertical density profile, and (c) distribution of horizontal velocities for several values of $\chi$ at $r_0/R\alpha= 0.63$ for the regular system size with $64$ squirmers and 	 $\alpha = 0.8$.
	Inset in b): distributions of vertical orientations.
	d) Mean swirling parameter $\langle W\rangle_\mathrm{cl}$ 
	(blue dots), averaged over all clusters and time, and mean cluster size $N_\mathrm{cl}$ (green triangles) plotted versus $\chi$.}
\label{fig:chi_hist}
\end{figure}

We again probe the order of the system by calculating the radial distribution functions for pushers, pullers, and rotlet-dipole squirmers. 
They are shown in Figs.~\ref{fig:beta_hist} and \ref{fig:chi_hist} together with density profiles. We also provide videos of the puller 
squirmers for $\beta=2$ at torques $r_0/R\alpha=0.32$ (M5) and $r_0/R\alpha=0.63$ (M6) as well as the rotlet-dipole squirmers 
with $\chi=1.0$ (M7).

\paragraph{$B_2$ mode}
Pusher and puller squirmers generate force-dipole flow fields through which they also interact with each other~\cite{BerkeLauga2008,SpagnolieLauga2012}. In particular, the additional flow vorticity 
presented in eq.~(\ref{eq:vort_fdp})
competes with the external and other hydrodynamic torques and thus contributes to the gyrotactic mechanism. As a consequence, the squirmer type $\beta$ has a profound influence on the radial distribution function $g(r)$, as Figs.~\ref{fig:beta_hist} a) and b)
demonstrate. At the lower torque value the peak
corresponding to the mean cluster distance 
for neutral squirmers has disappeared completely for $\beta = \pm 2$ and clustering does not exist.
The vertical density profiles in Fig.~\ref{fig:beta_hist} c) underline this: 
pushers and pullers reach larger heights while neutral squirmers are concentrated close to the botton since their
clusters cause collective sinking. 
In accordance with these findings, 
video M5 for puller squirmers with $\beta=2$ does not show clustering.

Doubling the external torque [Figs.~\ref{fig:beta_hist} b) and d)] still does not lead to visible clustering for the pusher system. 
This agrees with previous findings that cluster formation
in pusher systems is generally 
not favored~\cite{EvansLauga2011,AlarconPagonabarraga2013,RuehleStark2020}. However, the behavior of the
puller system ($\beta = 2$) 
changes drastically for the higher torque. First, we now see a weak peak
at a distance $8R$ in the radial distribution function
in Fig.~\ref{fig:beta_hist} b) indicating the existence of 
squirmer clusters. Indeed, we also observe small clusters in video M6. 
This is in agreement with the density profile in Fig.~\ref{fig:beta_hist} d), which is now more concentrated close to the 
bottom wall due to the collective sinking in clusters in contrast to the smaller torque value. The density profile also has two 
pronounced peaks, one of which is directly at the bottom wall. 
They indicate the stacked configurations of clusters typically observed for neutral squirmers already at smaller torques.
Interestingly, in $g(r)$ an additional, pronounced peak at a distance below the nearest-neighbor peak at $r=2R$ has developed,
which we discuss further below.

The puller clusters have a different structure compared to the neutral squirmer clusters. Video M6 shows pullers in a cluster 
typically pointing towards the cluster center. This agrees with the typical hydrodynamic puller-puller attraction along the swimmer axis and mutual reorientations towards each other due to the force-dipole vorticity
~\cite{LaugaPowers2009}. 
The tilt of the puller squirmer towards the horizontal is also caused by its near field close to a no-slip wall~\cite{IshikawaPedley2006}. 
The pronounced tilt of the puller axis away from the vertical is also visible in the distribution of vertical orientations in the  inset of Fig.~\ref{fig:beta_hist} d). Even a weak local maximum at $\cos\vartheta < 1$ appears.

Like neutral squirmers, puller clusters develop a stacked configuration.
The second peak in the density profile of the pullers
in Fig.~\ref{fig:beta_hist} d) belongs to a second layer of squirmers on top of the bottom layer. 
Note that the inwards pointing squirmers in the bottom layer induce fluid
flow towards the cluster center such that the squirmer from the second layer settles there. 
Thus, the cluster has a pyramidal three-dimensional structure. Assuming a pyramid with an equiliateral triangle 
as the base gives a horizontal distance 
of $\sqrt{3}/2 R$ between the nearest squirmers from 
the upper and lower layers and a vertical distance of $3/2 R$. We indicate these values as vertical dashed lines in Figs.~\ref{fig:beta_hist} b) and d), respectively.  They agree well with the peaks found 
from the simulations.

\paragraph{$C_2$ mode}
\label{sec:cmode}

In Fig.~\ref{fig:chi_hist} a)-c) we respectively
show radial distribution functions, vertical density profiles, and distributions of horizontal velocities for neutral squirmers ($\beta=0$) 
and several rotlet-dipole parameters $\chi$. 
Clusters form for all values except $\chi=3.3$, where no peak is visible in the radial distribution function beyond the nearest-neighbor 
distance. Furthermore, 
only the vertical distribution of squirmers with $\chi=3.3$ extends to larger heights, whereas it is 
concentrated close to the bottom for the other $\chi$ values, typical for squirmer clusters. Thus, strong 
rotlet-dipole flow prevents cluster formation. 
In addition, Fig.~\ref{fig:chi_hist} c) shows that with increasing $\chi$ the horizontal velocity components increase,
which also contributes to the disappearance of clusters at high $\chi$.
This is also indicated in Fig.\ \ref{fig:chi_hist} d), which plots
$\langle N_\mathrm{cl}\rangle$ versus $\chi$ (green  triangles).

Indeed, an inspection of video M7 reveals that already at $\chi=1.0$ squirmers stay more loosely bound to each other and clusters are less compact. Furthermore, we observe that
squirmers in a cluster swirl around the cluster center 
in anti-clockwise direction.
In eq.~(\ref{eq:rdp_z}) we show the angular velocity,
which a rotlet dipole close to a no-slip wall experiences due to the image flow. It is known to induce circular trajectories~\cite{IshimotoGaffney2013}, for example, of \emph{E.coli} bacteria~\cite{DiLuzioWhitesides2005,LaugaStone2006}. 
While \emph{E.coli} swim in circles with their orientation vector parallel to the wall, our squirmers have a strong vertical bias. 
According to eq.~(\ref{eq:rdp_z}) this 
corresponds to an angular velocity with reversed sign, as long as $\cos\vartheta$ exceeds $\sqrt{1/3}\approx 0.58$. This is indeed the case, as the inset in Fig.~\ref{fig:chi_hist} b) shows. The constant in-plane reorientation combined with the squirmers' self-propulsion results in circular swimming trajectories and in our simulations creates the swirling clusters.

We introduce the swirling parameter $W$ for a given cluster with a cluster center at $\overline{\mathbf{r}}$
by averaging the angular velocity of the circling
squirmers in the cluster:
\begin{equation}
W = \left\langle \frac{ \left(\mathbf{r} - \mathbf{\overline{r}}\right)\times \mathbf{v}}{\vert\mathbf{r} - \mathbf{\overline{r}} \vert^2} \cdot \mathbf{e}_z\right\rangle \, .
\end{equation}
We only take clusters with a cluster size $N_\mathrm{cl} > 2$ into account. In Fig.~\ref{fig:chi_hist} d) we show this quantity averaged 
over all clusters and $2.5\cdot 10^5$ timesteps, normalized by the inverse ballistic time scale $v_0/R$.
Clearly, at $\chi=0$ the squirmers do not swirl. For non-zero $\chi$ the normalized swirling parameter jumps to a value
with magnitude around $0.15 v_0/R$, and afterwards fluctuates in a narrow region between $0.15 v_0/R$ and $0.2v_0/R$.
For $|\chi|=3.3$, squirmers group together for short times, even though no long-term stable clusters emerge. Therefore we can still measure a value of $W$.

Since the chirality of the squirmer system changes with the sign of $\chi$, we can induce clock-wise motion with $\chi< 0$, corresponding to $\langle W\rangle <0$. 
We show this case in video M8, where $\chi=-1$.

\subsubsection{Influence of gravity}
\label{sec:gravity}

\begin{figure}
\centering
\resizebox{0.45\textwidth}{!}{\includegraphics{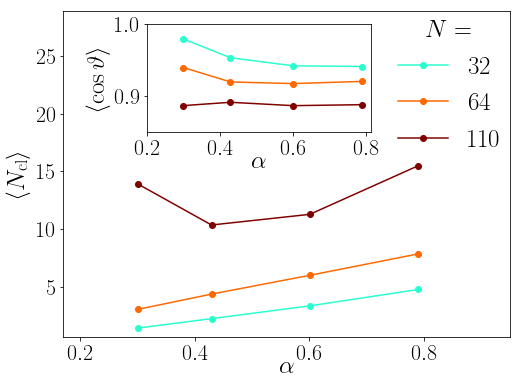}}

\caption{Mean cluster size of neutral squirmers plotted versus $\alpha$ for several values of total squirmer 
	number $N$ at $r_0/R\alpha=0.32$.
	Inset: 
	Mean vertical orientation of squirmers plotted versus	$\alpha$.}
\label{fig:mean_cluster_size_alpha}
\end{figure}

The parameter $\alpha = v_0 / v_\mathrm{sed}$ determines how strongly squir\-mers are confined to the bottom wall.
On the one hand, gravity induces the stokeslet flow via which squirmers also interact with each other hydrodynamically. 
Thus, at smal\-ler $\alpha$ one expects squirmers to be more tilted against the vertical due to the stronger flow
vorticity. On the other hand, at stronger gravity squirmers are closer to the bottom wall and therefore the hydrodynamic interaction
with the no-slip wall also forces a stronger upright orientation in addition to the gravitational torque, see eq.~(\ref{eq:rdp_phi})
\cite{SpagnolieLauga2012,RuehleStark2018}. Thus, the behavior of the system is determined by these two effects.

In Fig.~\ref{fig:mean_cluster_size_alpha} we plot the mean cluster size $\langle N_\mathrm{cl} \rangle$
versus $\alpha$. It increases monotonically for $N=32$ and $N=64$. Thus, clusters become smaller at higher gravity and 
finally disintegrate into single squirmers. In order to find the reason for this behavior, we measured the orientational distributions 
and show the mean vertical orientation in the inset.
The general trend is that for decreasing $\alpha$ (increasing gravity), the squirmer orientation is more aligned with the vertical. Thus, the stronger vertical alignment of neutral squirmers
due to their hydrodynamic interactions with the wall is able to counterbalance the action of the stronger stokeslet vorticities. This results in more compact and smaller clusters, similar to the effect of larger 
gravitational torques in Sect.~\ref{sec:cluster_size}.

For the system with $N=110$ squirmers the mean cluster size increases again for the lowest value $\alpha=0.3$. Due to the stronger confinement to the bottom wall
and the larger areal density, the squirmers are in close contact to each other which increases the measured cluster size. Furthermore, the mean vertical orientation does not increase further at $\alpha=0.3$ in contrast to the smaller squirmer numbers.
Due to the larger density, the mutual reorientation due to flow vorticity is stronger and counters the 
vertical alignment due to hydrodynamic interactions with the bottom wall.

We note that as the clusters become smaller, they also start to rotate as a whole. In particular, we observe rotating
squirmer trimers at large gravity. We mention them since they resemble the spinner states found at very high gravity in lattice-Boltzmann simulations~\cite{ShenLintuvuori2019}. 
In these studies the vertical alignment of neutral squirmes is solely due to their hydrodynamic interactions with the wall [see eq.\ (\ref{eq:rdp_phi})]. Furthermore, in the limit of very large gravity ($\alpha \rightarrow 0$), our squirmer system reaches the regime of Ref.~\cite{KuhrStark2019}, where at zero external torque a quasi-hexagonal monolayer  (Wigner fluid) of squirmers with upright orientation is formed.

\section{Conclusions}
\label{sec:conclusions}

We have investigated the cluster formation
of bottom-heavy squirmers under strong gravity 
floating above the bottom surface. Already a squirmer pair mimicking the minuet dance, found in 
experiments with the Volvox algae, reveals the basic gyrotactic mechganism \cite{DrescherGoldstein2009,IshikawaGoldstein2020}.
While the stokeslet vorticity from the squirmer neighbor tilts the squirmer towards the neighbor so that they swim
towards each other, the gravitational torque stabilizes the upright orientation. This enables horizontal oscillations of the squirmer positions.

At higher squirmer numbers, the vorticity-induced mutual attraction leads to the formation of squirmer clusters, however only for sufficiently large gravitational torques.
We quantify the cluster properties by the two-dimensional density distribution, the mean cluster size, and the radial distribution function, from which we extract the mean cluster distance. Our simulations find that horizontal motion 
within the clusters decreases with increasing torque
and the clusters become more compact. The cluster size has a maximum at a finite torque
that is close but below the value where the vorticity-induced rotation of touching neighbors can be balanced. 
The  maximum standard deviation of the cluster size 
at the same torque value shows the volatility of the clusters in the emerging cluster state. 
Increasing the torque further, squirmers in the clusters are less mobile. 
The clusters become more compact  and ultimately 
the cluster size reaches a constant value for each density. The mean distance between the clusters 
also decreases with increasing torque and saturates.
Interestingly, the areal density has little effect on the mean cluster distance. 

Furthermore, we investigated how force and rotlet di\-poles infuence cluster formation. Puller squirmers require larger torques in order to form clusters, because their additional flow-field vorticity adds to the effective attraction and thus the stabilizing gravitational torque has to be larger. Their clusters are situated closer to the bottom wall where hydrodynamic wall interactions tilt squirmers towards each other.
In contrast, pushers do not form noticeable clusters in our simulations.
Squirmers with a rotlet dipole also form clusters for small dipolar strengths and due to their interaction with the no-slip wall
perform swirling motion. 
Finally, for increasing gravitational strength squirmers move closer to the botton wall. The hydrodynamic  wall interactions contribute to the vertical alignment of neutral squirmers, which also decreases the cluster size.
In conclusion, the swimmer hydrodynamics 
has a profound effect on the cluster formation
when the occuring hydroynamic multipoles besides the stokeslet also contribute to the flow vorticity, which has to be balanced by the gravitational torque.

Gyrotactic cluster formation could be realized for micro\-swim\-mers, such as Janus particles or Volvox algae, which are indeed
bottom-heavy~\cite{SimmchenSanchez2016,DrescherGoldstein2009}.
Janus particles also offer the possibility to tune the strength of the gravitational torque,
while for Volvox algae it exhibits a natural variation. For an L-shaped particle the torque was controlled by changing the relative arm lengths and thus the effective lever arm~\cite{tenHagenBechinger2014}.
However, the realization of the external torque is not decisive for our results. For example, phototaxis is an alternative way of inducing orientational alignment and the strength of the vertical reorientation can be controlled by light
intensity~\cite{WilliamsBees2011,FragkopoulosBaeumchen2021}. 
But here, the size and orientation of the 
part of the swimmer, which is illuminated, also influences the torque, as was shown for a phototactic Janus particle~\cite{SinghFischer2018}. Microswimmers casting shadows on each other, influence their collective dynamics.
Furthermore, gradients in oxygen concentration also orient bacteria
along the vertical, which was observed for \emph{B. subtilis} swimming upwards towards 
an air-liquid interface at the top~\cite{SokolovAranson2009}.
However, in general, any chemotaxis towards chemical fields, such as nutrients, can profoundly affect the dynamics of microswimmers as exemplified by the effect of trail avoidance in 
active emulsion droplets~\cite{JinMaass2017}. These droplets also form floating, spontaneously rotating clusters, when the 
concentration of the surfactant fuel is high enough~\cite{HokmabadPHD}.
Thus, while we have identified the generic hydrodynamic
mechanism of gyrotactic cluster formation, the influence of chemical fields needs further study.

Finally, we mention two possible extensions of our investigations. Biological 
microswimmers typically move in non-Newtonian fluids \cite{RileyLauga2014,ZoettlYeomans2019}. 
Therefore, it is worthwile to investigate, how robust gyrotactic cluster formation is for squirmers in complex fluids. 
This requires to extent the method of MPCD to viscoelastic media \cite{GompperWinkler2009}.
Furthermore, current research also explores the dynamics of non-spherical swimmers~\cite{Roberts2010,SenguptaStocker2017,HeckelSimmchen2021}, including squirmer rods~\cite{TheersWinkler2016Soft,ZantopStark2020}. The collective motion of rod-shaped particles is strongly influenced by their steric interactions. Under a gravitational torque, this suggests that the vertical orientation is even more stable in a dense suspension with smectic order, because collisions with neighboring rods counteract tilting against the vertical. Therefore, the interesting question arises how this affects the formation of gyrotactic clusters.

\section{Authors contributions}
All the authors were involved in the preparation of the manuscript.
All the authors have read and approved the final manuscript.

\section*{Acknowledgement}
We thank B.V. Hokmabad, C.C. Maass and J. Grawitter for fruitful discussions. This project was funded by Deutsche Forschungsgemeinschaft through the priority program SPP1726 (grant number STA352/11). The authors acknowledge the 
North-German Supercomputing Alliance (HLRN) for providing HPC resources that have contributed to the research results reported 
in this paper.

\bibliographystyle{epj}
\bibliography{shorttitles,lit}
%
%

\end{document}